%% file: ms.tex
\documentclass[iop,apj]{emulateapj}
\slugcomment{Accepted for publication in Astrophysical Journal Letters}

\usepackage[]{amsmath}
\usepackage[dvipsnames]{xcolor}
\usepackage{bm}
\usepackage{ulem}
\shorttitle{A Passive-side of GLASS: Metal deficient galaxies}
\shortauthors{Morishita et al.}

\usepackage{color}		
\definecolor{midgray}{gray}{0.4}	
\definecolor{orange}{rgb}{1,0.5,0} 
\definecolor{blue}{rgb}{0,0,0.6}  
\usepackage{mathptmx}
\usepackage{times}

\definecolor{ao}{rgb}{0.0, 0.0, 1.0}
\usepackage[colorlinks=true, citecolor=ao, filecolor=ao, linkcolor=ao, urlcolor=ao]{hyperref}     
\usepackage{scrextend}
\deffootnote[0.5em]{0em}{1em}{\textsuperscript{\thefootnotemark}\,}
\input{ctx}

\newcommand{\myemail}{tmorishita@stsci.edu}
\newcommand{\simgt}{\,\rlap{\lower 3.5 pt \hbox{$\mathchar \sim$}} \raise
1pt \hbox {$>$}\,}
\newcommand{\simlt}{\,\rlap{\lower 3.5 pt \hbox{$\mathchar \sim$}} \raise
1pt \hbox {$<$}\,}
\newcommand{\Msun}{M_{\odot}}
\newcommand{\Mstel}{M_{\ast}}
\newcommand{\logm}{\log M_*/\Msun}
\newcommand{\logZ}{\log Z_*/Z_\odot}

\newcommand{\kms}{{\rm km~s^{-1}}}
\newcommand{\oii}{[\textrm{O}~\textsc{ii}]}

\def\Nph{19} % Sample which satisfy photo-z and magnitude.
\def\Nex{2} % Sample which have spectra without significant contamination.

\newcommand{\hst}{{\it HST}}

\newcommand{\spit}{{\it Spitzer}}

\newcommand{\Z}{{\rm [Z/H]}}
\newcommand{\FH}{{\rm [Fe/H]}}

\newcommand{\alp}{{\rm [\alpha/Fe]}}

\newcommand{\sext}{{SExtractor}}

\newcommand{\griz}{{\ttfamily Grizli}}

\newcommand{\resp}{respectively}

\newcommand{\affilA}{Space Telescope Science Institute, 3700 San Martin Drive, Baltimore, MD 21218, USA}
\newcommand{\affilB}{Department of Physics and Astronomy, UCLA, 430 Portola Plaza, Los Angeles, CA 90095-1547, USA}
\newcommand{\affilC}{Department of Astronomy, University of California, Berkeley, CA 94720-3411, USA}
\newcommand{\affilD}{School of Physics and Astronomy, University of Minnesota,116 Church Street SE, Minneapolis, MN 55455, USA}
\newcommand{\affilE}{University of California Davis, 1 Shields Avenue, Davis, CA 95616, USA}
\newcommand{\affilF}{Leibniz-Institut f\"ur Astrophysik Potsdam, An der Sternwarte 16, 14482 Potsdam, Germany}
\newcommand{\affilG}{School of Physics, Tin Alley, University of Melbourne VIC 3010, Australia}

\begin{document}
\title{
Metal Deficiency in Two Massive Dead Galaxies at $\bm{z\sim2}$
}

\author{
T.~Morishita\altaffilmark{1},
L.~E.~Abramson\altaffilmark{2}, 
T.~Treu\altaffilmark{2},
X.~Wang\altaffilmark{2},
G.~B.~Brammer\altaffilmark{1},
P.~Kelly\altaffilmark{3,4},
M.~Stiavelli\altaffilmark{1},
T.~Jones\altaffilmark{5},
K.~B.~Schmidt\altaffilmark{6},
M.~Trenti\altaffilmark{7},
B.~Vulcani\altaffilmark{7}
}
\affil{$^1$\affilA; \href{mailto:\myemail}{\myemail}}
\affil{$^2$\affilB}
\affil{$^3$\affilC}
\affil{$^4$\affilD}
\affil{$^5$\affilE}
\affil{$^6$\affilF}
\affil{$^7$\affilG}

% ======================================================================
\begin{abstract}

Local massive early-type galaxies are believed to have completed most of their star formation $\sim10$\,Gyr ago and evolved without having substantial star formation since.
If so, their progenitors should have roughly solar stellar metallicities ($Z_*$), comparable to their values today.
We report the discovery of two lensed massive ($\logm\sim11$), $z\sim2.2$ dead galaxies, that appear markedly metal deficient given this scenario.
Using 17-band {\it HST}+$K_{s}$+{\it Spitzer} photometry and deep $\hst$ grism spectra from the GLASS and SN Refsdal follow-up campaigns covering features near $\lambda_{\rm rest}\sim4000$\,\AA, we find these systems to be dominated by A-type stars with $\logZ=-0.40\pm0.02$ and $-0.49\pm0.03$ ($30$-$40\%$ solar) under standard assumptions. The second system's lower metallicity is robust to isochrone changes, though this choice can drive the first system's from $\logZ=-0.6$ to 0.1.
If these two galaxies are representative of larger samples, this finding suggests that evolutionary paths other than dry minor-merging are required for these massive galaxies.
Future analyses with direct metallicity measurements---e.g., by the {\it James Webb Space Telescope}---will provide critical insight into the nature of such phenomena.

\end{abstract}

\keywords{galaxies: evolution -- galaxies: abundances -- galaxies: structure}

% =====================================================================

\section{Introduction}
\label{sec:intro}

Stellar chemical compositions are a powerful means to link galaxies across cosmic time and therefore understand their evolution.
Massive early-type galaxies in the local universe exhibit solar or super-solar stellar metallicities \citep[$Z_*$; e.g.,][]{thomas05}.
This holds beyond $z\sim0$---previous observations report similar values in dead massive galaxies up to $z\sim1.6$ \citep[e.g.,][]{gallazzi14, onodera15, lonoce15}, indicating that either they formed from metal-enriched gas or retained a lot of their metals after forming from pristine gas.

These observations spanning $\sim10$\,Gyr seem to fit in a recently popular evolution scenario for massive galaxies, where dry minor merging might grow galaxy sizes without adding much mass \citep[e.g.,][]{naab09}, explaining the observed size evolution by a factor of $\simgt3$ from $z\sim2$ to 0 \citep[e.g,][]{szomoru12, vanderwel14}.
As these galaxies are already dead and near the mass of local ellipticals \citep[e.g.,][]{stiavelli99, daddi05, glazebrook17}, they are believed to evolve without further substantial episodes of in situ star formation over the rest of their lives.
If so, the main $z\sim2$ progenitor must already have {\it at least} its $z=0$ descendant's metallicity: the low-mass satellites it will accrete are probably metal poor \citep{gallazzi05, kirby13}, and thus can only dilute the final system's total metallicity \citep[see also][]{choi14}.

The above scenario, however, may not be the complete picture.
Other works \citep[e.g.,][]{newman12, nipoti12, sonnenfeld14} show that minor mergers alone cannot accomplish the necessary size evolution. 
Furthermore, recent stellar metallicity estimates at $z\simgt2$ yield values $\sim$2.5$\,\sigma$ below the local $M_{*}$--$Z_{*}$ locus \citep{kriek16, toft17}, adding to arguments for more-than-minor-merging growth.
To clarify the picture, we must take steps towards a more complete  galaxy census.

Here we report two additional massive ($\logm\sim11$), dead galaxies at $z\sim2.2$ that appear substantially under-enriched. 
Both galaxies have typical A-type spectra, but $\logZ=-0.40\pm0.02$ and $-0.49\pm0.03$, respectively. 
With the other spectrophotometric results at $z\sim2$, these galaxies suggest something other than dry minor merging is required to explain the subsequent evolution of some fraction of $z\simgt2$ passive galaxies.
While results based on a larger sample will be presented in a future study, this Letter reports the discovery of these two galaxies and briefly discusses its implications.
Throughout, magnitudes are quoted in the AB system assuming $\Omega_m=0.3$, $\Omega_\Lambda=0.7$, $H_0=70\,\kms\, {\rm Mpc}^{-1}$.

%%%%%%%%%%%%%%%%%%%%%
\begin{figure*}
\centering
	\includegraphics[width=0.98\textwidth]{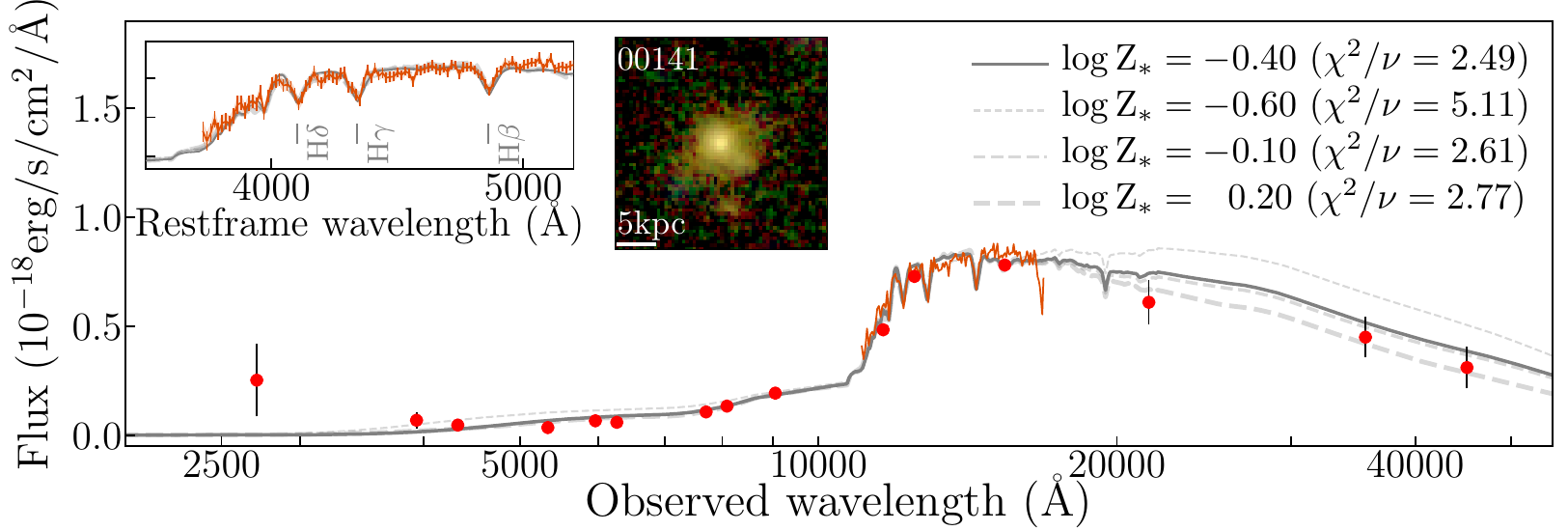}
	\includegraphics[width=0.98\textwidth]{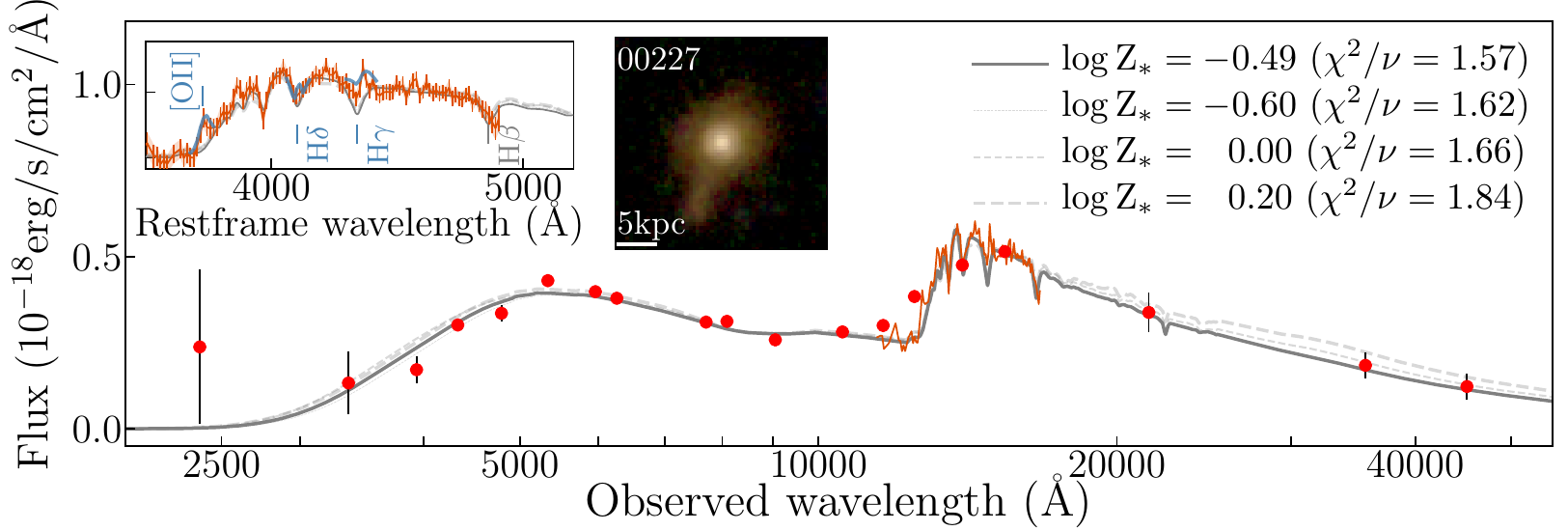}
	\caption{
	SEDs and fits for 00141 (top) and 00227 (bottom).
	Photometry and spectra (red points and lines) are shown with errors (black and red bars, respectively) overlaid on the best fit model (gray solid line).
	Gray dashed lines show fit models with different metallicities at local $\chi^2/\nu$-minima ($\logZ\sim[-0.6, 0, 0.2]$: thin to thick).
	Insets show the region around $\lambda_{\rm rest}\sim4000$\,\AA\ and $3\!\farcs6\times3\!\farcs6$ F814/125/160W pseudo-color images of the sources.
	Weak emission lines are detected in 00227 and modeled with gaussian (blue lines).
	}
\label{fig:spec}
\end{figure*}

%%%%%%%%%%%%%%%%%

\section{Data and Sample}
\label{sec:data}

\subsection{\hst\ Data}
We require high-quality spectral coverage over the rest frame 3700--4200\,\AA\ window and broadband photometry over a much wider wavelength range to pin-down details of the youngest and oldest stellar populations and break the age--metallicity degeneracy. 
Such data are difficult to obtain from the ground, but lensing magnification and deep NIR \hst\ spectrophotometric campaigns provide an ideal way to study passive galaxies at such early epochs.

The spectral range of the WFC3/G141 grism covers age sensitive features---the Balmer and $4000$\,\AA\ breaks---for galaxies at $z\sim2$, such that high-quality \hst\ photometry covering $0.2$--$1.7\,\mu{\rm m}$ from CLASH \citep{postman12} and HFF \citep{lotz17} can constrain stellar metallicities at sufficient confidence.
Among the survey lines of sight, the $z=0.544$ galaxy cluster MACSJ1149.5+2223 is an optimal target: it is covered by 17-band photometry spanning F225W--F160W to typical $5\,\sigma$ depths of $\sim29$ (for the HFF data) to $\sim27$\,mag (for CLASH). 
Moreover, spectroscopy from GLASS \citep{schmidt14, treu15} and the follow-up HSTGO/DDT campaign of supernova Refsdal \citep[Proposal ID 14041;][]{kelly15, kelly16} provide 34\,orbit-deep G141 spectra at 4\,position angles (PAs; 81000\,s total integration). 

We match all imaging to F160W resolution using unsaturated stars as PSFs before conducting photometry with \sext\ \citep{bertin96}.
Apertures are set to radii of 0.\!\arcsec72, but measurements are scaled based on the aperture-to-total flux ratio in F160W.

As for spectra, we retrieve the WFC3/G141 data from the Mikulski Archive for Space Telescopes archive, and run the \griz\footnote{\url{https://github.com/gbrammer/grizli}} software package for one-dimensional optimal extraction.
We then stack the extracted spectrum from each orbit using inverse-variance weighting based on the pipeline-output uncertainties.
We use jackknife resampling to estimate the error on the stack's mean \citep[e.g.,][]{onodera15} and add that to the $\sigma_{\rm RMS}$ formal error to obtain our final spectral uncertainties.
We reach $S/N\sim20$ per $\sim$45\,\AA\ pixel at rest-frame 4000\,\AA.
This is slightly finer than the Nyquist sampling of G141 thanks to sufficient dithering across the orbits, each of which produces a shift in the dispersion direction.

The line spread function (LSF) is estimated using each galaxy's morphology at each PA, to which the fitting templates are convolved (Section~\ref{sec:sed}).
While the spectra and templates are matched to the broadest one among all PAs, LSF variations at different PAs show negligible variation given the sources' point-like/symmetric morphology.

%%%%%%%%%%%%%%%%%%%%%%
\subsection{K$_{s}$ and IRAC Photometry}
\label{ssec:ksimg}
Beside precise spectroscopic constraints near $\lambda_{\rm RF}\sim4000$\,\AA, well-characterized spectral energy distribution (SED) features at rest-frame $\simgt1\,\mu{\rm m}$ are essential for estimating dust and metallicity content.
This regime is redward of \hst's bandpasses for $z\simgt1$ galaxies, but accessible via ground-based $K_s$ and \spit/IRAC photometry. 
We obtain $K_s$ and IRAC1 (3.6\,$\mu{\rm m}$)/IRAC2 (4.5\,$\mu{\rm m}$) photometry from the publicly available catalog by \citet{castellano16}, based on Keck/MOSFIRE \citep[][]{brammer16} and Spitzer Frontier Fields observations (P.I.~T.~Soifer, P.~Capak).

To account for aperture-loss differences compared to the \hst\ photometry, as well as zeropoint uncertainties in the ground-based data, we renormalize the public data based on corrections derived from the best-fit \hst-only SED templates for $z<1$ galaxies, where \hst\ photometry captures the full wavelength range to $\lambda_{\rm RF}\sim1\,\mu$m.
The medians of this correction are $-2.8\%$, $-3.0$\%, and $3.4\%$ for $K_s$, IRAC1, and IRAC2, respectively. 
We incorporate these small corrections into the photometric error budget.

%%%%%%%%%%%%%%%%%%
\subsection{Sample Galaxies}
\label{ssec:sample}
To cover the 4000\,\AA\ break at sufficient $S/N$, we limit our initial pre-selection to galaxies with $H_{160}<24$ and $1.6<z_{\rm phot}<3.3$.
A rest-frame color cut at $U-V>1.0$ is then applied to bias the sample towards passive galaxies. 
Combined, these criteria leave \Nph\,suitable galaxies in the MACS1149 field. 
Unfortunately, contamination from nearby spectra affects most of these, leaving a final sample of \Nex\, galaxies with sufficiently clean spectra.
The spectra have sufficient $S/N$ of $\sim20$ per pixel at rest-frame 4000\,\AA\ ($B\sim21.7$ and 22.4\,mag, respectively).
The rest of the sample will not support robust results in the following SED analysis.

%%%%%%%%%%%%%%%%%%%%%%%%%%
\begin{deluxetable*}{lrcccccccccccc}
\tabletypesize{\footnotesize}
\tablecolumns{14}
\tablewidth{0pt} 
\tablecaption{
Summary of the SED fitting parameters. 
}
\tablehead{\colhead{ID} & \colhead{RA} & \colhead{DEC} & \colhead{$z_{\rm spec.}$} & \colhead{$\mu$} & \colhead{$\log Z_*$} & \colhead{$a_0$} & \colhead{$a_1$} & \colhead{$a_2$} & \colhead{$a_3$} & \colhead{$a_4$} & \colhead{$A_V$} & \colhead{$M_*$}\\
\colhead{} & \colhead{(degree)} & \colhead{(degree)} & \colhead{} & \colhead{} & \colhead{($Z_\odot$)} & \colhead{[10\,Myr]} & \colhead{[100\,Myr]} & \colhead{[500\,Myr]} & \colhead{[1\,Gyr]} & \colhead{[3\,Gyr]} & \colhead{(mag)} & \colhead{($10^{10}M_\odot$)}}
\startdata
\cutinhead{Padova isochrone}
00141 & 177.403 & 22.419 & $1.967_{-0.001}^{+0.002}$ & $1.85_{-0.11}^{+0.83}$ & $-0.40_{-0.02}^{+0.02}$ & $0.18_{-0.13}^{+0.26}$ & $0.22_{-0.16}^{+0.40}$ & $74.06_{-4.85}^{+2.86}$ & $1.47_{-1.09}^{+2.07}$ & $0.52_{-0.38}^{+0.73}$ & $1.79_{-0.03}^{+0.02}$ & $17.86_{-0.54}^{+0.43}$\\
00227 & 177.407 & 22.416 & $2.412_{-0.004}^{+0.004}$ & $1.68_{-0.09}^{+0.37}$ & $-0.49_{-0.03}^{+0.04}$ & $0.71_{-0.62}^{+1.77}$ & $14.72_{-1.89}^{+1.81}$ & $14.10_{-1.52}^{+1.59}$ & $0.88_{-0.67}^{+1.31}$ & $1.37_{-0.94}^{+1.22}$ & $0.66_{-0.07}^{+0.06}$ & $5.78_{-0.56}^{+0.68}$\\
\cutinhead{MIST isochrone}
00141 & 177.403 & 22.419 & $1.967_{-0.002}^{+0.001}$ & $1.85_{-0.11}^{+0.83}$ & $0.10_{-0.07}^{+0.07}$ & $0.23_{-0.17}^{+0.34}$ & $0.31_{-0.23}^{+0.52}$ & $42.78_{-5.59}^{+5.36}$ & $2.79_{-2.00}^{+3.28}$ & $0.94_{-0.70}^{+1.19}$ & $1.37_{-0.08}^{+0.08}$ & $14.72_{-1.14}^{+1.22}$\\
00227 & 177.407 & 22.416 & $2.411_{-0.002}^{+0.004}$ & $1.68_{-0.09}^{+0.37}$ & $-0.50_{-0.06}^{+0.07}$ & $0.67_{-0.51}^{+1.10}$ & $11.98_{-2.13}^{+1.80}$ & $17.75_{-1.31}^{+1.09}$ & $0.01_{-0.01}^{+0.05}$ & $0.40_{-0.31}^{+0.62}$ & $0.62_{-0.08}^{+0.06}$ & $5.39_{-0.40}^{+0.46}$
\enddata
\tablecomments{
50th percentile is quoted as the best fit parameter with 16$^{\rm th}$/84$^{\rm th}$ percentiles as statistical uncertainties.
Systematic uncertainties derived from simulations (Section \ref{sec:result}) are $\sim0.25$\,dex, $0.25$\,dex and $0.17$\,dex for $A_V$, $\logZ$, and $\logm$, respectively.
$a_i$: Amplitudes of the templates, with $i=[0,1,2,3,4]$ representing $t_i=[0.01, 0.1, 0.5, 1, 3]/$\,Gyr.
$\mu$: Median of magnification by the foreground cluster. 
Magnification errors are not included in the parameter errors as all are $\mu$-independent except for $\Mstel$.
}
\label{tab:res}
\end{deluxetable*}

%################
\begin{figure*}
\centering
	\includegraphics[width=0.49\textwidth]{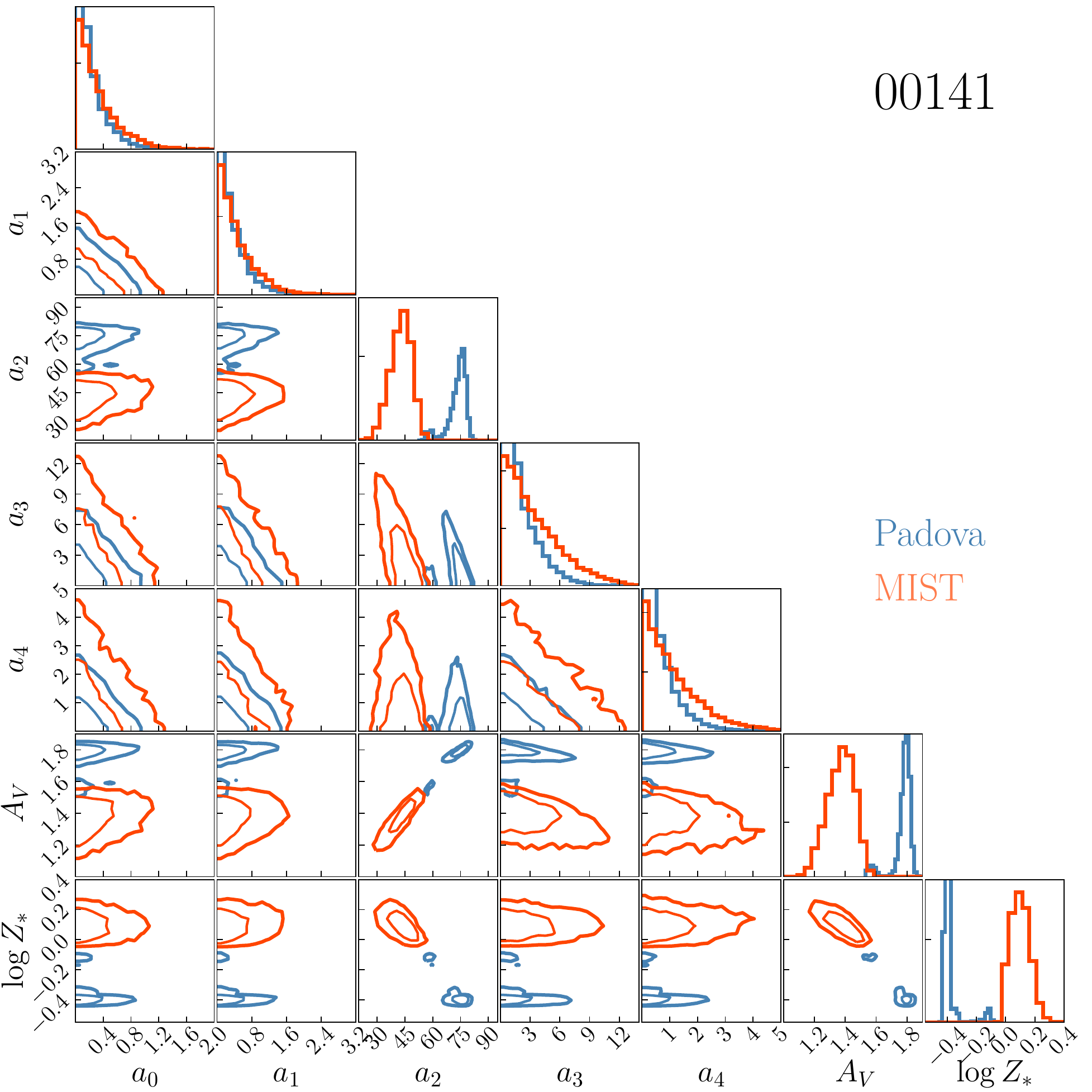}
	\includegraphics[width=0.49\textwidth]{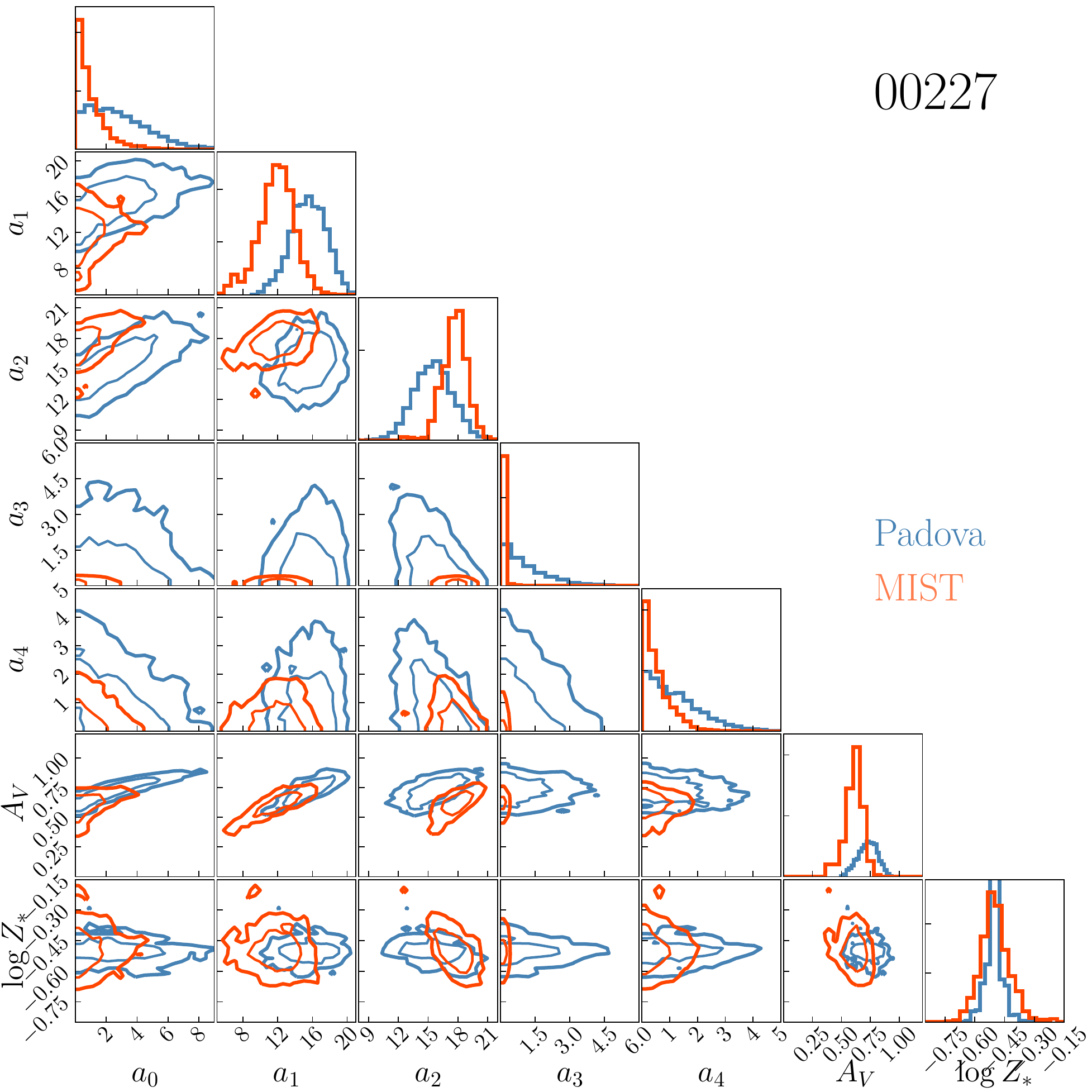}
	\caption{
	Covariance matrices of the SED fitting parameters, for 00141 (left) and 00227 (right), based on Padova (blue) and MIST (red) isochrones.
	Contours enclose the top 68\%, 95\%, and 99.7\% of the distribution, respectively.
	}
\label{fig:cov}
\end{figure*}
%################

%---------------------------------------------------------------------------------------------------------------------
\section{Spectro-photometric SED fitting}
\label{sec:sed}

With extensive multi-band photometry and deep NIR spectra, we deconstruct the stellar populations of the two high-$z$ passive galaxies. 
A forthcoming paper presents full details of the SED fitting, but our scheme was established and tested by \citet{kelson14}  using a similar data set at $z\lesssim1$.
The core of the method is to find the best combination of amplitudes, $\{a_{i=[0,1,2,3,4]}\}$, for a set of stellar templates of different ages, $\{t_{i=[0,1,2,3,4]}\}$, that matches the data. 
This flexibility avoids systematic biases in, e.g., stellar mass that can arise when using functional forms for star formation histories.

Our baseline results adopt the Flexible Stellar Population Synthesis code \citep[FSPS;][]{conroy09fsps, conroy10} to generate four simple stellar population (SSP) templates with ages of [0.01, 0.1, 0.5, 1.0, 3.0]\,Gyr (roughly O, B, early/late-A, and F stars), based on Padova isochrones and the MILES stellar library.
We assume a Salpeter IMF and Calzetti dust law. 
Our main results are robust to those assumptions except for the choice of isochrones, which systematically affects one system (see below).
The number and choice of templates was set to fully exploit the data based on result reproducibility tests using simulations.
Assuming non-zero durations of star formation for each template---e.g., exponentially declining star formation histories---pushes systems to older ages, and thus {\it lower} metallicities to compensate for the subsequently redder SEDs.
Thus, SSPs produce conservative results for our purposes.

The interpolation scheme in the FSPS python module generates each template at metallicities spanning $\logZ\in[-1.0, 0.5]$ in increments of 0.05\,dex. 
Metallicity and extinction are fixed across all five SSP templates; i.e., the fits have $5+1+1$ free parameters.

We use a Monte Carlo Markov Chain to find the best combination of $a_i$, $A_V$, $Z_*$, and their covariances in the parameter ranges ($a_i\in[0,200]$ and $A_V\in[0,4]$\,mag) with flat priors.
We set the number of walkers to 200 and use $N_{\rm mc}=10000$ realizations.
Redshifts determined beforehand via $\chi^2$-minimization (and visual inspection of absorption lines) are fixed during parameter estimation. Final results are drawn from the last $N_{\rm mc}/2$ steps to avoid biases from the initial exploration.
Emission lines detected at $>$1.5\,$\sigma$ across 3\,pixels ($\sim$135\,\AA) are masked from the best fit template of each iteration and later fit with gaussians.

One caveat is that the metallicity determined above is not a direct measurement, as metallicity sensitive indicators, such as Fe and Mg, fall outside the G141 spectral range.
The broadband-defined ``$Z_*$'' here is thus different from the total metallicity derived from absorption line measurements (see Section~\ref{sec:disc}).

%---------------------------------------------------------------------------------------------------------------------
\section{Results}
\label{sec:result}

Figure~\ref{fig:spec} shows the data and fits and Table~\ref{tab:res} summarizes the best-fit parameters.
Both galaxies prefer sub-solar metallicities under the standard assumption of Padova isochrones: $\logZ=-0.40_{-0.02}^{+0.02}$ and $-0.49_{-0.03}^{+0.04}$ for 00141 and 00227, respectively, with estimates and uncertainties reflecting the 16$^{\rm th}$/50$^{\rm th}$/84$^{\rm th}$ percentile values.\footnote{While it is known that metallicities are effectively differentiated in the rest-frame NIR \citep[e.g.,][]{lee07}, our $K_S$/IRAC data do not provide the tightest constraints. Instead, the grism spectrum and broadband photometry at rest-frame $\sim$2000---5000\,\AA\ do by breaking the age/metallicity degeneracy.}
Also, both best-fit templates are dominated by $\sim0.5$\,Gyr old stellar populations: $\sim99\%$ of 00141's light is in the $a_2$-template, while 00227's stellar population is younger, with $\simgt44\%$ in $a_1$ and $a_2$.
Stellar masses are $M_*=1.8\times10^{11}\,\Msun$ and $5.8\times10^{10}\,\Msun$, \resp.
00141 exhibits no significant emission, while 00227 shows very weak Balmer infilling (EWs of H$\delta\sim0.8\pm$\,0.2\,\AA\ and H$\gamma\sim1.8\pm$\,0.2\,\AA) and \oii\ emission ($\sim3.4\pm$\,0.7\,\AA), suggesting the latter was indeed quenched more recently. 
Both galaxies have compact/centrally concentrated morphologies with effective radii $r_e=1.8$--2\,kpc and S\'ersic indices of $n=8$.
These values are typical for passive galaxies at similar redshifts \cite[e.g.,][]{szomoru12}, and confirm that significant dry minor merging is needed if they are to grow to match the size of $z=0$ early-types.

Figure~\ref{fig:cov} shows the fit parameter covariances.
While some covary strongly (e.g., $a_2$ and $A_V$), none significantly impact the metallicity assessment.
Simulations of our SEDs (discussed momentarily) confirm these results. 

As we lack detailed spectra, the metallicities reported here might depend on the isochrones adopted during SED fitting. 
We tested all isochrones available in FSPS, and found that, while 00227's result is robust, $Z_{*}$ changes significantly for 00141, reaching a low of -0.6 \citep[PARSEC;][]{bressan12} and a high of $\logZ\sim0.10\pm{0.07}$ \citep[MIST;][]{choi16}.
This is mostly caused by the different treatments of the asymptotic giant branch, which contributes significantly to the galaxy's luminosity at ages near 00141's of $\sim500$\,Myr. Since we cannot determine the best isochrone with the current data set, we base our discussion on the result with the standard Padova isochrone to facilitate contextualizing our results with those in the literature. However, we present the MIST results in Table~\ref{tab:res} and related figures to illustrate the effect of this choice.

In addition to isochrone systematics, there is a chance that, due to dust or statistical fluctuations, an observed sub-solar SED could be measured from a galaxy that is {\it truly} enriched to solar levels.
To estimate this contamination rate, we tested our fitting method on 1000 simulated SEDs covering the same spectrophotometric bandpasses at comparable $S/N$ with parameters randomly chosen over the fitting range.
While full results will be discussed in our forthcoming study, the simulation shows that the input and extracted estimates are unbiased up to $A_{V}=1.3$\,mag.
Above that, our inferences {\it underestimate} $A_{V}$ by up to 0.2\,mag. 
If so, to return a sufficiently reddened SED, the fitter would tend to {\it increase} $Z_{*}$, returning an {\it overestimate} of the true metallicity, making our results conservative. 
Finally, we find $\sim$20\% and $\sim$5\% chances that 00141's $\logZ=-0.4$ and 00227's $\logZ=-0.49$, respectively, could have come from a galaxy with truly solar metallicity.

%################
\begin{figure}
\centering
	\includegraphics[width=0.46\textwidth]{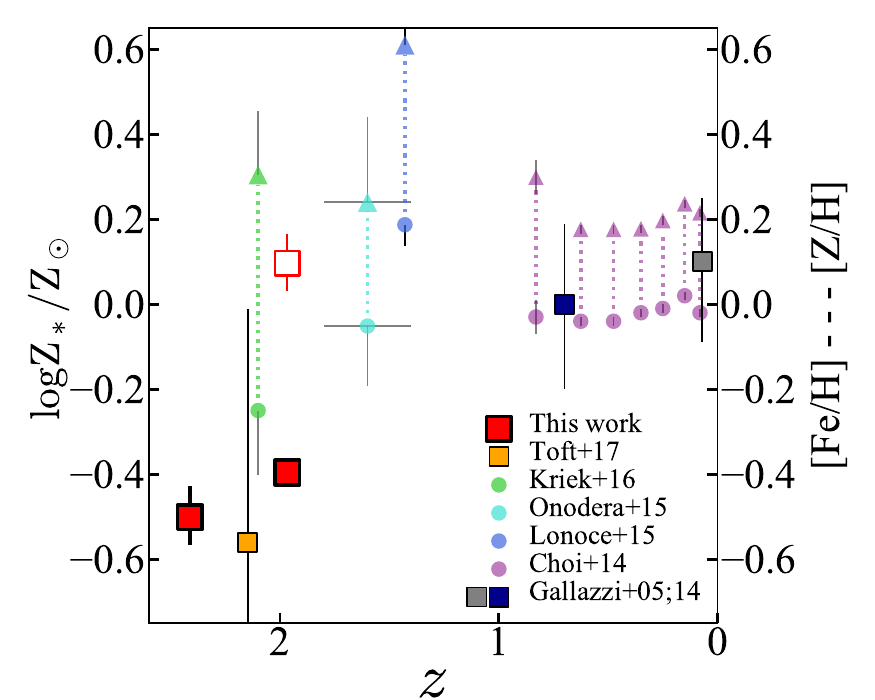}
	\caption{
	 Stellar metallicities over time comparing our results using Padova (red filled squares) and MIST isochrones (open squares) to previous literature $\logZ$ estimates \citep[][small squares]{gallazzi05,gallazzi14,toft17} of massive galaxies ($\logm\simgt11$).
	We show \FH\ and \Z\  (circles and triangles, respectively) estimated from direct iron/$\alpha$-abundances \citep[][]{choi14, onodera15, lonoce15, kriek16}, though the direct comparison should be made with care (see main text).
	}
\label{fig:zZ}
\end{figure}
%################

\section{Implications of Sub-solar Metallicities in Massive Dead Galaxies\label{ssec:subZ}}
\label{sec:disc}

The sub-solar abundances found in this study, suggest that these galaxies are in the early stages of {\it chemical} evolution, if already quite mature in terms of {\it mass}.
While perhaps not surprising given the age of the universe (e.g., \citealt{delucia07}), these results do differ from the solar to super-solar values reported previously for massive quiescent galaxies at lower redshifts \citep[][]{gallazzi05, gallazzi14, choi14, onodera15} as summarized in Figure~\ref{fig:zZ}. 
(\citealt{gallazzi14} and \citealt{leethochawalit18}, however, show $Z_{*}$ evolution at $z<0.7$.)

Now, $Z_{*}$ here derives from broad SED features. Given the complexities of stellar evolution, it is therefore perhaps not easily interpreted, e.g., as reflecting iron or total (iron + $\alpha$) abundances. This fact complicates comparisons to other measurements based on highly detailed spectroscopy. As such, in those cases \citep{choi14,lonoce15,onodera15,kriek16} we show in Figure~\ref{fig:zZ} both \FH\ and the {\it total} abundance, \Z, where $\Z = \FH + 0.94 \alp$ \citep{thomas03}. Other results reported only as ``$\logZ$'' are reproduced as published \citep{gallazzi05,gallazzi14,toft17}.

Assuming our two galaxies do not represent an outlier population, the differences illustrated in Figure \ref{fig:zZ} raise interesting questions about the diversity of evolutionary paths of these objects to $z\sim0$.
What can be securely said is that the popular scenario for high-$z$ passive galaxy evolution does not hold for these objects: while it may grow their sizes sufficiently, dry minor merging cannot correspondingly raise their metallicities to those of their $z=0$ counterparts. 
Given their low mass, any satellites are almost certain to have lower $Z_{*}$ than these centrals \citep{pasquali10, kirby13}, so such merging would only further dilute these systems' metal content.
Furthermore, most metallicity gradients in local early-types are negative---or at least flat---with higher metallicity in their centers \citep[e.g.,][]{martin-navarro18}. 
If the objects studied here were to somehow merge with more metal-rich satellites, their $z=0$ gradients would go in the opposite direction \citep[][]{nipoti12}.

Given this boundary condition, there appear to be three plausible routes for our galaxies to take:

{\bf 1) Major mergers:} Merging with another $\sim$equal-mass galaxy of $Z_{*}\gtrsim{\rm Z_{\odot}}$ would bring them towards/onto the local mass-metallicity relation. 
While major mergers are rare, globally, and probably too rare to happen to {\it all} massive galaxies \citep[e.g.,][]{bundy09}, given the amount of time from the observed redshift to today and the increasing merger rate increases towards higher-$z$ \citep[e.g.,][]{lotz11}, this route cannot be excluded here.
If this is taken as the preferred solution, measurements like this may provide independent tests of the merger rate once sufficient samples exist.

{\bf 2) A second burst of star formation.}
Newly accreted gas delivered either by {\it wet} satellite galaxies or cosmic accretion can  raise metallicities at later times via {\it in situ} star formation. 
Dissipation efficiently supplies the gas to the central part of the system and induces starbursts. 
Despite uncertainties in the metallicity of the inflowing gas and the size of these bursts---perhaps $\sim10$-$15\%$ of the systems' extant stellar mass \citep[e.g.,][]{macarthur08}---the large numbers of minor mergers in massive systems since $z\sim2$ \citep[e.g.,][]{rodriguez15} make this scenario attractive, as do observations of blue core early-types \citep[e.g.,][]{treu05}, and other metrics of minor merger-triggered star formation at intermediate redshift \citep[e.g.,][]{kaviraj11}.
The centrally concentrated nature of star formation in this scenario would also produce negative metallicity gradients, qualitatively consistent with those in the local universe.
However, the increase of the central density would result in smaller galaxy size, contradicting the observed size evolution.

{\bf 3) The galaxies remain as they are.}
If this is the case, their descendent systems must be outliers today, suggesting most of the size {\it and} metallicity evolution of the passive {\it population} is due purely to progenitor bias. 
The situation is unclear at present: \citet{poggianti12} find local compact galaxies (size outliers) to have younger light-weighted ages, which could also reflect low metallicities. Yet, other studies find about solar values for compact local relics \citep[e.g.,][]{martin-navarro18}.
More spectroscopic studies of local compact galaxies will help resolve this picture further.

\section{Summary}

We identify two galaxies whose SEDs imply substantially sub-solar stellar metallicities under standard assumptions, adding to recent, similar results.
Taken together, these findings suggest that some---if not many---$z\simgt2$ passive galaxies are unlikely to evolve exclusively through dry minor merging (a currently popular scenario).
As \citet{kriek16} also argue, this might imply that $z\simgt2$ galaxies are a distinct population from those even at $z\sim1.5$, which already exhibit solar metallicities (see also \citealt{newman12}).
Rather, the observed metallicity evolution of massive galaxies could be driven by newly quenched galaxies \citep[a.k.a.\ progenitor bias;][]{vandokkum96, carollo13}, as may also partially explain their size growth.

Ground based $K_s$ and IRAC photometry covers the rest-NIR in this study, a regime key to robust metallicity measurements \citep[e.g.,][]{lee07}. 
Despite the depth of such data, our lack of high-resolution spectroscopy does leave systematic uncertainties that must wait to be resolved by JWST. Most notable among these are isochrone effects, which can have a substantial impact at ages $\sim$500\,Myr. Once in-hand, large samples of direct JWST iron and $\alpha$-element abundances at high-$z$ would bear perhaps on our understanding of the cosmic merger rate, and the duty cycle of late time star formation in once-dead galaxies in addition to their early chemical evolution.

\acknowledgements
We thank the anonymous referee for insightful suggestions, and Camilla Pacifici and Mia Bovill for fruitful discussion.
Support for GLASS (HST-GO-13459, 14041) was provided by NASA through a grant from the Space Telescope Science Institute, operated by the Association of Universities for Research in Astronomy, Inc., under contract NAS 5-26555.
This work uses gravitational lensing models by PIs Brada{\v c}, Natarajan \& Kneib (CATS v4.1), Merten \& Zitrin, Sharon, Williams, Keeton, Bernstein \& Diego, and the GLAFIC group.
B.V. acknowledges support from an Australian Research Council Discovery Early Career Researcher Award (PD0028506).

{\it Software:} Astropy \citep{muna16}, EMCEE \citep{foreman13}, FSPS \citep{conroy09fsps, conroy10, foreman14}, Grizli (G.~Brammer), lmfit \citep{newville17}.

%==================================================================

\end{document}

%% file: ctx.tex
\usepackage{etoolbox}
\makeatletter
\patchcmd{\NAT@citex}
  {\@citea\NAT@hyper@{\NAT@nmfmt{\NAT@nm}\NAT@date}}
  {\@citea\NAT@nmfmt{\NAT@nm}\NAT@hyper@{\NAT@date}}
  {}% Do nothing if patch works
  {}% Do nothing if patch fails
\patchcmd{\NAT@citex}
  {\@citea\NAT@hyper@{%
     \NAT@nmfmt{\NAT@nm}%
     \hyper@natlinkbreak{\NAT@aysep\NAT@spacechar}{\@citeb\@extra@b@citeb}%
     \NAT@date}}
  {\@citea\NAT@nmfmt{\NAT@nm}%
   \NAT@aysep\NAT@spacechar%
   \NAT@hyper@{\NAT@date}}
  {}% Do nothing if patch works
  {}% Do nothing if patch fails
\patchcmd{\NAT@citex}
  {\@citea\NAT@hyper@{%
     \NAT@nmfmt{\NAT@nm}%
     \hyper@natlinkbreak{\NAT@spacechar\NAT@@open\if*#1*\else#1\NAT@spacechar\fi}%
       {\@citeb\@extra@b@citeb}%
     \NAT@date}}
  {\@citea\NAT@nmfmt{\NAT@nm}%
   \NAT@spacechar\NAT@@open\if*#1*\else#1\NAT@spacechar\fi%
   \NAT@hyper@{\NAT@date}}
  {}% Do nothing if patch works
  {}% Do nothing if patch fails
\makeatother